\documentclass[prx,twocolumn,superscriptaddress,nofootinbib,floatfix]{revtex4-2}
\usepackage{soul}
\usepackage{amsfonts}
\usepackage{graphicx}
\usepackage{bm}
\usepackage{amssymb}
\usepackage{color}
\usepackage{amsmath}
\usepackage{amstext}
\usepackage{latexsym}
\usepackage{enumitem}
\usepackage[usenames,dvipsnames]{xcolor}
\usepackage[colorlinks=true,citecolor=Cerulean,linkcolor=RubineRed,urlcolor=Cerulean]{hyperref}

\newcommand\ba{\begin{eqnarray}}
\newcommand\ea{\end{eqnarray}}
\newcommand\be{\begin{equation}}
\newcommand\ee{\end{equation}}

\newcommand{\bra}[1]{\langle #1|}
\newcommand{\ket}[1]{|#1\rangle}

\usepackage{braket}
\usepackage{float}

\def\De{\Delta}

\usepackage{verbatim}
\usepackage{graphicx}
\begin{document}
\author{Lakshmija N K}
\email{222414004@smail.iitpkd.ac.in}
 \affiliation{Department of Physics, Indian Institute of  Technology Palakkad, Kerala, 678623.}

 \author{Uma Divakaran}
\email{uma@iitpkd.ac.in}
 \affiliation{Department of Physics, Indian Institute of  Technology Palakkad, Kerala, 678623.}

\title{Optimal performance in a chaotic Floquet Quantum Battery}

\date{\today}

\begin{abstract}
We study a quantum battery consisting of $N$ qubits interacting uniformly with each other. Starting from its ground state, the battery is charged by introducing periodic kicks in the transverse direction along with adding disorder in the interaction direction.   Although the battery Hamiltonian along with the initial state preserves the permutation symmetry, the dynamics in presence of disorder during the charging phase takes the system out of the permutation symmetric subspace, resulting to a larger amount of stored energy as compared to that without disorder. In the limit of large disorder, this interplay brings in a possibility of a chaotic charging phase when stored energy can be exactly calculated using random matrix theory. More importantly, we also find that the energy fluctuations decrease and become negligible in the limit of large disorder, which is also when the energy stored is maximum, a condition highly desirable for the optimum functioning of the battery. {We complete the battery cycle by considering the storage phase in the presence of dephasing noise. We find that for larger disorder strengths, the ergotropy remains close to the stored energy, indicating nearly complete energy extractability with an efficiency approaching unity.}
%We complete the battery cycle by looking at the storage phase in presence of a dephasing noise, and comment on the extraction of energy by studying the ergotropy .
\end{abstract}


\maketitle
%%%%%%%%%%%%%%%%%%%%%%%%%%%%%%%%%%%%%%%%%%%%%%%%%%%%%%%%%%%%%%%%%%%%%

%\textcolor{blue}{Replace $t_Q$ by $\tau$}\\

\section{Introduction}
The research on quantum batteries has increasingly focused on harnessing collective quantum phenomena for enhanced charging and energy storage. A quantum battery consists of a quantum system in which energy is stored in the Hamiltonian via unitary \cite{alicki2013entanglement} or non-unitary dynamics \cite{NonunitarySantos}. In contrast to its classical counterparts, quantum batteries have the advantage of exploiting inherently quantum phenomena like collective interactions, entanglement, and coherence, to improve the performance of the battery\cite{alicki2013entanglement, PhysRevResearch.2.023113, binder2015quantacell, andolina2018charger, PhysRevB.99.035421, PhysRevLett.120.117702, PhysRevLett.125.236402, PhysRevA.97.022106, PhysRevA.101.032115, PhysRevLett.118.150601, PhysRevB.99.205437, PhysRevB.100.115142, PhysRevLett.129.130602, RevModPhys.96.031001}. This has prompted substantial theoretical and experimental research in this field in the recent years \cite{hymas2025experimental, YANG2024102300, PhysRevA.106.042601, PhysRevB.108.L180301, PhysRevA.109.062614, g45c-ssfx, Hu2026} . 

One of the major challenges in this setting is to identify charging protocols and many body dynamics that can simultaneously optimize the stored energy, charging power and the stability of the charged state, in order to design a robust battery. Several protocols have been proposed to address these requirements. Initial studies focused on collective charging schemes, in which multiple battery cells are charged collectively through global interactions or a common charging medium, showing enhanced charging power compared to parallel charging\cite{PhysRevLett.118.150601,PhysRevLett.120.117702,binder2015quantacell}. Similarly, intrinsic spin-spin interactions in many-body batteries can enhance the charging power, particularly when the interactions are sufficiently long-ranged\cite{PhysRevA.97.022106}. However, an improvement in charging power alone does not necessarily guarantee a large stored energy or a stable charged state. Also, various charging protocols that yield the same average stored energy can differ significantly in their energy fluctuations and charging precision\cite{Friis2018}. In open-system settings, stable charging can be achieved through different mechanisms, like engineered dissipation leading to stable active steady states \cite{Barra2019}, and adiabatic protocols suppressing spontaneous discharging \cite{NonunitarySantos}. Moreover, dark-state protocols can enhance energy storage while stabilizing the charged state through quantum interference\cite{Quach2020}, while non-Markovian memory effects can enable complete charging and help preserve the stored energy over longer times \cite{Kamin2020}. At the same time, stability can also arise directly from the underlying many-body dynamics, like in fast-scrambling Sachdev-Ye-Kitaev batteries, which exhibit strongly suppressed temporal fluctuations of the average stored energy, producing a stable and precise charged state\cite{Rosa2020}. Therefore, identifying dynamical regimes which provide a favorable balance among these properties remain an important problem. {An additional important criterion for assessing battery performance is the extractable work, quantified by the ergotropy, since not all of the stored energy is necessarily available for extraction\cite{Allahverdyan2004, PhysRevLett.122.047702,PhysRevLett.129.130602, watanabe2026scalinglawasymptoticfreedom}.}

In this regard, periodically driven Floquet systems\cite{PhysRev.138.B979,Sambe1973,sato2025floquet,Bukov2015,Goldman2014,Eckardt2017} offer a versatile platform for implementing charging protocols involving controlled energy injection via repeated driving cycles\cite{PhysRevE.105.044125, puri2024floquet, romero2025kicked, PhysRevA.102.060201, mazumdar2025sunburst,Chen2020,Sreeram2025,Romero2026Dissipative,Mazumdar2026,Shukla2026}. Mondal and
Bhattacharjee\cite{PhysRevE.105.044125} showed that resonant driving enhances the stored energy and stabilizes the charged state in a periodically driven transverse-field Ising spin chain battery. On the other hand, Puri et al.\cite{puri2024floquet} demonstrated that the combined effect of periodic driving and long-range
interactions leads to super-extensive scaling of the charging power, providing a genuine
quantum advantage in a Floquet-driven long-range interacting spin model. In a subsequent work, Romero et al.\cite{romero2025kicked} showed that appropriately designed Floquet kick sequences enable fast, robust, and nearly optimal charging using a periodically driven kicked Ising spin chain. In the case of open-systems, periodic charging and discharging cycles can generate Floquet bound states that suppress decoherence and improve the stability of quantum batteries\cite{PhysRevA.102.060201}. In separate studies not related to batteries, it has been shown that the long time dynamics in chaotic Floquet systems is often well described by random-matrix-theory \cite{d2016quantum,Trail2008,Haake1999,Olsacher2022,Chan2018,Ray2016}, for which kicked top is a well studied model \cite{haake1987classical, fox1994chaos, wang2004entanglement, PhysRevA.78.042318, chaudhury2009quantum, PhysRevE.98.052205,haake1991quantum}. Introducing disorder in the kicked top  lifts the existing permutation symmetry and allows the dynamics to extend from the permutation-symmetric subspace to the full Hilbert space\cite{lakshminarayan2025chaos,8kgb-c2xb}.
%At the same time, introducing disorder in a many body Floquet system gives rise to interesting Physics\cite{lazarides2015fate,Ponte2015Ergodic,Bordia2017,Haldar2017,Wauters2019,Zhang2022}. For instance, the presence of disorder in the kicked top(a textbook example of quantum chaos)\cite{haake1987classical, fox1994chaos, wang2004entanglement, PhysRevA.78.042318, chaudhury2009quantum, PhysRevE.98.052205,haake1991quantum} lifts the existing permutation symmetry and allows the dynamics to extend from the permutation-symmetric subspace to the full Hilbert space\cite{lakshminarayan2025chaos,8kgb-c2xb}. 
In another work, the authors show the existence of many body localized phase and delocalized phase by tuning the periodicity of one of the Hamiltonians in a Floquet system\cite{ponte2015many} .

While several studies focused on utilizing collective effects and coherent driving in the functioning of the quantum battery \cite{alicki2013entanglement,PhysRevResearch.2.023113, PhysRevLett.120.117702,PhysRevLett.118.150601,PhysRevB.99.205437}, the influence of dynamical features such as chaos, ergodicity, and symmetry breaking on the preformance of the battery has received comparatively less attention \cite{PhysRevB.100.115142,Sreeram2025,Romero2025Scrambling,Ho2026LMGBattery}, which we try to address in this work .

For this, we investigate an all-to-all interacting battery system which is charged with a Floquet Hamiltonian. We focus on studying the effect of introducing disorder in the periodically driven charging phase in terms of any improvement in the performance of the battery.
Our calculations indicate a dynamical crossover that separates a strongly fluctuating regime from a reliable operating regime characterized by increased energy storage, reduced energy fluctuations and enhanced extractability, thereby providing a clear operational advantage of improved battery stability when the strength of the disorder is relatively large.

The paper is structured as follows: In Sec. \ref{sec:model}, we introduce the kicked top model and its characteristic features relevant to our study. We discuss the battery Hamiltonian and the adopted charging protocol in Sec. \ref{sec:charging} followed by a discussion on the metrics  used to characterize battery performance in Sec.\ref{sec:metrics}. Additionally, Sec. \ref{sec:storage} focuses on the storage and extraction stage of the battery. The main results of our work is discussed in Sec \ref{sec:results}. Sec. \ref{sec:summary} summarises our findings and outlines directions for future work.

\section{THE KICKED TOP MODEL}
\label{sec:model}
In this section, we discuss the kicked top model, which is associated with the charging phase of the proposed battery.
The kicked top is a prototypical model of quantum chaos widely used as a standard testbed for exploring ergodicity and universal Random Matrix Theory(RMT) behavior in driven quantum systems \cite{haake1987classical, fox1994chaos, wang2004entanglement, PhysRevA.78.042318, chaudhury2009quantum, PhysRevE.98.052205, haake1991quantum}. The fully connected Hamiltonian describing $N$ spin-$\frac{1}{2}$ particles along with periodic kicks can be written as:
\begin{equation}
 H=\frac{k}{4N }\sum_{l,l^{'}=1}^{N}\sigma_{l}^{x}\sigma_{l^{'}}^{x}  +\frac{p}{2}\sum_{n=-\infty }^{\infty }\sum_{l=1}^{N}\sigma_{l}^{y}\delta \left(\frac{t}{\tau}-n\right),
\label{eq_spin}   
\end{equation}
where $\sigma_l^r (r=x,y,z)$ represent Pauli spin operators acting on the site $l$ with $k$ as the strength of all-to-all spin interactions along $x$ direction, $p$ is the angle of turn per kick (or kick strength) in the $y$-direction and $\tau$ is the time interval between two successive kicks. The parameter $k$, also known as ``chaoticity parameter", can be tuned to study the transition from regular (small $k$) to fully chaotic dynamics (large $k$). The system possesses a permutation symmetry under exchange of spins.  Consequently, the above Hamiltonian can be formulated in terms of the collective angular momentum operators $J_r=\sum_{l=1}^N\sigma_l^r/2$ where $r=x,y,z$ as:
\begin{equation}
H=\frac{k}{N }J_{x}^2 + p J_{y}\sum_{n=-\infty }^{\infty }\delta \left(\frac{t}{\tau}-n\right).
\label{eq_kicked}
\end{equation}
The total angular momentum is conserved since $[H,J^2]=0$, restricting the dynamics within a fixed quantum number $j$. It is usually chosen to be the maximum symmetric subspace corresponding to $j={N}/{2}$ fixed by the initial state chosen, as discussed later,  restricting the Hilbert space to $(N+1)$ dimensions. The Floquet operator ($\hbar=1$ throughout the paper) governing the stroboscopic evolution over one driving period $\tau$ is defined as \cite{PhysRevA.78.042318, haake1987classical}:
\begin{equation}
U_0= \exp\left({-i\frac{k}{N}J_{x}^{2}\tau} \right) \hspace{0.1 cm} \exp\left({-i pJ_{y}\tau}\right).
\label{eq_floquet_cl}
\end{equation}

The model admits a well-defined classical limit as $j \rightarrow \infty$ (in the thermodynamic limit). It is well established that in the kicked top classical phase space, regular dynamics dominate for  $k<2$, whereas for $k>6$, the dynamics is fully chaotic.  The intermediate $k$ regime shows a mixed phase space\cite{PhysRevA.78.042318,8kgb-c2xb}.

The permutation symmetry of the model can be broken by introducing disorder in the interaction term of the Hamiltonian, as demonstrated in \cite{lakshminarayan2025chaos}. The authors show that in the limit of large disorder, the dynamics explore the full Hilbert space  where the expectation values of various observables are given by RMT corresponding to the full Hilbert space. We use a similar disordered version of the model in the charging of the battery, which is given by:
\begin{equation}
H=\frac{k}{2N }\sum_{l<l^{'}}(1+\epsilon_{ll^{'}})\sigma_{l}^{x}\sigma_{l^{'}}^{x}  +\frac{p}{2}\sum_{n=-\infty }^{\infty }\sum_{l=1}^{N}\sigma_{l}^{y}\delta \left(\frac{t}{\tau}-n\right),
\label{eq_spin_disorder}
\end{equation}
where $\epsilon_{ll^{'}}$ term introduces a quenched disorder, and is taken randomly from a Gaussian distribution with zero mean and standard deviation $w$, also called disorder strength. The corresponding Floquet operator $U_w$ is of the form:
\begin{equation}
    U_{w}=\exp\left(-i\frac{k\tau}{2N}\sum_{l<l^{'}}(1+\epsilon_{ll^{'}})\sigma_{l}^{x}\sigma_{l^{'}}^{x}\right) \hspace{0.1 cm} \exp\left(-i\frac{p\tau}{2}\sum_{l=1}^{N}\sigma_{l}^{y}\right).
    \label{dynamics1}
\end{equation}
As discussed earlier, in the disorder free case corresponding to $w=0$, the dynamics can be analyzed within the permutation symmetric subspace (PSS) having dimension $(N+1)$. However, in the disordered case with $w \neq 0$, the dynamics is pushed out of PSS. For large disorder, the dynamics will be chaotic, spanning the full Hilbert space (FHS) of $2^N-$dimensions \cite{lakshminarayan2025chaos}. We employ the above dynamics to charge the battery as discussed in the next section.

\section{THE BATTERY CYCLE}
\label{sec:battery_protocol}
In this section, we present the protocol used for charging along with the metrics used to analyze battery performance, followed by storage and extraction. %\lnk{(see Fig. \ref{battery cycle}).}

\subsection{Charging protocol}\label{sec:charging}

While most of the existing studies focus on batteries composed of non-interacting qubits\cite{PhysRevLett.125.236402,PhysRevB.100.115142,PhysRevLett.129.130602,mazumdar2025sunburst,puri2024floquet,romero2025kicked,watanabe2026scalinglawasymptoticfreedom}, we instead consider a battery consisting of $N-$ spin half particles with all-to-all interactions given by a Hamiltonian
\begin{equation}
    H_{B}=-\frac{k}{4N}\sum_{l,l^{'}=1}^{N} \sigma^x_l \sigma^x_{l^{'}}.
    \label{battery}
\end{equation}
This choice allows us to explicitly demonstrate the enhancement of battery performance that arises from the breaking of permutation symmetry under the charging protocol considered here. 

Charging is implemented via a periodically driven Floquet protocol. Each charging cycle consists of addition of disorder in the all-to-all interacting spin chain, along with $n-$kicks in the transverse direction. In other words, the battery consisting of the Hamiltonian given by Eq. \ref{battery} is evolved up to n-kicks with unitary $U_w$ defined in the Eq. \ref{dynamics1}. 
To ensure consistency with the battery Hamiltonian, we introduce an overall negative sign in the interaction term of the evolution operator without affecting overall dynamics. 
We fix $\tau=1$ and $p=\frac{\pi}{2}$ throughout this paper. The battery is prepared in a spin-coherent state\cite{haake1991quantum,2015AmJPh..83...30L,PhysRevA.13.357} represented by 
\begin{equation}
    {\vert{\theta,\phi}\rangle}={\left(\cos{\frac{\theta}{2}}\vert{0}\rangle+e^{i\phi}\sin{\frac{\theta}{2}}\vert{1}\rangle\right)}^{{\otimes}N}.
    \label{scs}
\end{equation}
These states display minimum uncertainty for finite angular quantum number ``$j$", with the relative variance of the total angular momentum vanishing as $1/j$, showing classical behavior when $j\rightarrow\infty$. {The evolved state after $n$ Floquet periods is then given by $\ket{\psi(n)}=U_{w}^n\ket{\psi_0}$ where $\ket{\psi_0}$ is the initial spin coherent  state.}

We now discuss various metrics used to analyze the performance of the battery.

\subsection{BATTERY PERFORMANCE METRICS}
\label{sec:metrics}

We present below suitable metrics to quantitatively assess the charging dynamics of the battery. These metrics characterise the energy gained, energy transfer rate and the charged state's robustness under repeated driving.  
\begin{itemize}
    \item \textbf{Stored energy}: It determines the charging capacity after $n$ charging cycles, or quantifies how much energy is stored in the battery relative to its initial state, given by
    \begin{equation}
        E(n)= \bra{\psi(n)}{H_B}\ket{\psi(n)} - \bra{\psi_0}{H_B}\ket{\psi_0}.
    \end{equation}
    
    \item \textbf{Maximum stored energy:} It is defined as the maximum value of the stored energy attained during the charging dynamics:
    \begin{equation}
    E_{max}\equiv \max_n E(n).
    \end{equation}

    \item \textbf{Charging power}: It captures the rate of energy deposition given by
    \begin{equation}
        P(n)=\frac{E(n)}{n\tau}.
    \end{equation}

    Following the definition of the charging power, one can also define the maximum charging power as its maximum value attained during the charging process.

     \item \textbf{Energy Variance}: It measures fluctuations of the final energy determining the stability and reliability of the battery, and can be calculated as
     \begin{equation}
         (\De{E(n)})^2= \bra{\psi(n)}{H_B^2}\ket{\psi(n)} - (\bra{\psi(n)}{H_B}\ket{\psi(n)})^2.
     \end{equation}
\end{itemize}

In this work, we consider the disorder-averaged maximum stored energy, $\langle E_{\max}\rangle_{w}$, obtained by averaging the maximum stored energy over different realizations of the disorder.
Additionally, the disorder and long time averaged energy stored and its corresponding variance are denoted as $\overline{\langle{E(n)}\rangle_{w}}$ and $\overline{\langle{ (\Delta E(n))^2}\rangle_{w}}$, respectively, which are obtained after averaging over 100 disorder realizations and performing a long-time average over the interval ranging from 600 to 1000 kicks.

\begin{figure}[h]
\centering
\includegraphics[width=1.0\linewidth]%, height=0.8\linewidth]
{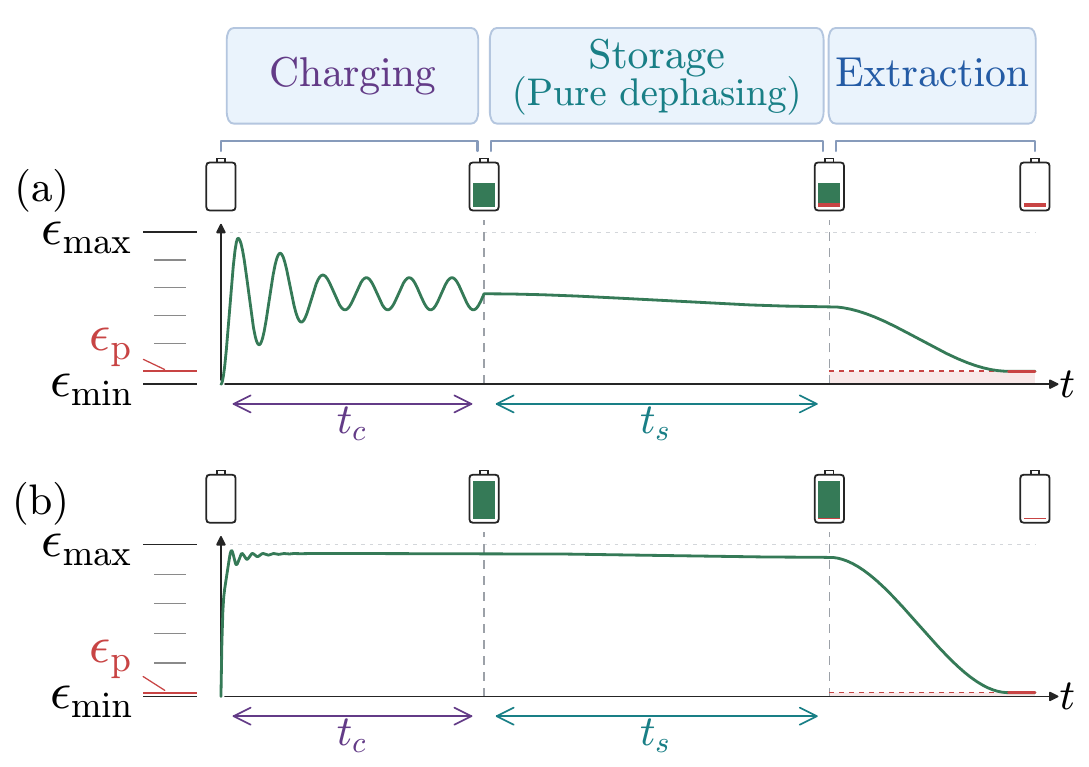}
\caption{\textit{Battery cycle}: Schematic illustration of charging, storage and work extraction for (a) small and (b) large disorder introduced during the charging stage. Here, green and red battery fillings represent extractable energy and passive energy, respectively. The latter is locked up in the battery and cannot be extracted as work through cyclic unitary operations. The levels $\epsilon_{min}$ and $\epsilon_{max}$ denote the minimum and maximum battery eigen energies, while $\epsilon_{\mathrm{p}}$ represent the final passive energy.  For small disorder, the extractable energy exhibits persistent oscillations during charging over a duration $t_c$ and decreases under pure dephasing during the subsequent storage interval $t_s$; for large disorder, it rapidly saturates near the full battery capacity during the charging phase, and undergoes only a very slight reduction during the storage.}
\label{battery cycle}
\end{figure}

\subsection{STORAGE AND EXTRACTION}
\label{sec:storage}

So far, we have focused on the charging of the battery. We now turn to the storage stage and the subsequent energy extraction. The periodic charging along with disorder is switched off after $n$-kicks, and the battery enters the storage stage. During this stage, it is usually subjected to a noisy channel. For convenience, we choose pure dephasing noise during the storage stage.

Pure dephasing is a form of decoherence that suppresses quantum coherence between different energy eigenstates while conserving the system’s energy (leaving all populations unchanged) \cite{Breuer2002}. In many experimentally relevant regimes, decoherence is predominantly governed by dephasing arising from low-frequency fluctuations in the environment and control parameters, whereas energy relaxation occurs over longer timescales \cite{Krantz2019}. Therefore, pure dephasing provides a minimal description of decoherence during idle storage without introducing energy exchange with environment. The corresponding master equation (see Appendix \ref{appen:noise} for details), formulated using the Lindblad operator $ \mathcal{L}= \sqrt{2\gamma} H_B$, is given by \cite{Shastri2025}:
\begin{equation} \label{lindblad}
    \frac{d \rho}{ d t}= -i[H_B, \rho]-\gamma [H_B, [H_B, \rho]].
\end{equation}
After the storage phase in which the battery evolves under the above equation, we now discuss the energy extraction process.
Since not all stored energy is extractable as useful work, ergotropy\cite{Allahverdyan2004,Pusz1978} provides a realistic measure of the battery's work-extraction capability. Ergotropy is the maximum amount of work that can be extracted from a quantum system by cyclic unitary operations, 
and is defined as:
\begin{equation}
    \mathcal{E}(\rho)=\operatorname{Tr}(\rho H_B)-\min_{\substack{U }}\operatorname{Tr}(U\rho U^\dagger H_B),
\end{equation}
where the minimization is performed over all possible unitary operators ($U$). The minimum energy attained through unitary transformations corresponds to the passive state $\rho_p$, which is obtained by rearranging the eigenvalues of $\rho$ in decreasing order onto the eigenstates of $H_B$, which are themselves ordered according to increasing energy. Consequently, the ergotropy can be rewritten as:
\begin{equation}
    \mathcal{E}(\rho)=\operatorname{Tr}(\rho H_B)-\operatorname{Tr}(\rho_p H_B).
\end{equation}

Following the extraction stage, the battery may be reinitialized to its prescribed initial state, thereby enabling repeated charging–storage–extraction cycles. In the present work, however, we restrict our attention to the performance of a single battery cycle and do not explicitly model the reinitialization process. The charging, storage, and work-extraction stages considered here are illustrated schematically in Fig.~\ref{battery cycle}, the details of which will be clear when we discuss the numerics in the next section.

\section{Numerical RESULTS}
\label{sec:results}

\begin{figure}[h]
\centering

\includegraphics[width=1.0\linewidth]%, height=0.8\linewidth]
{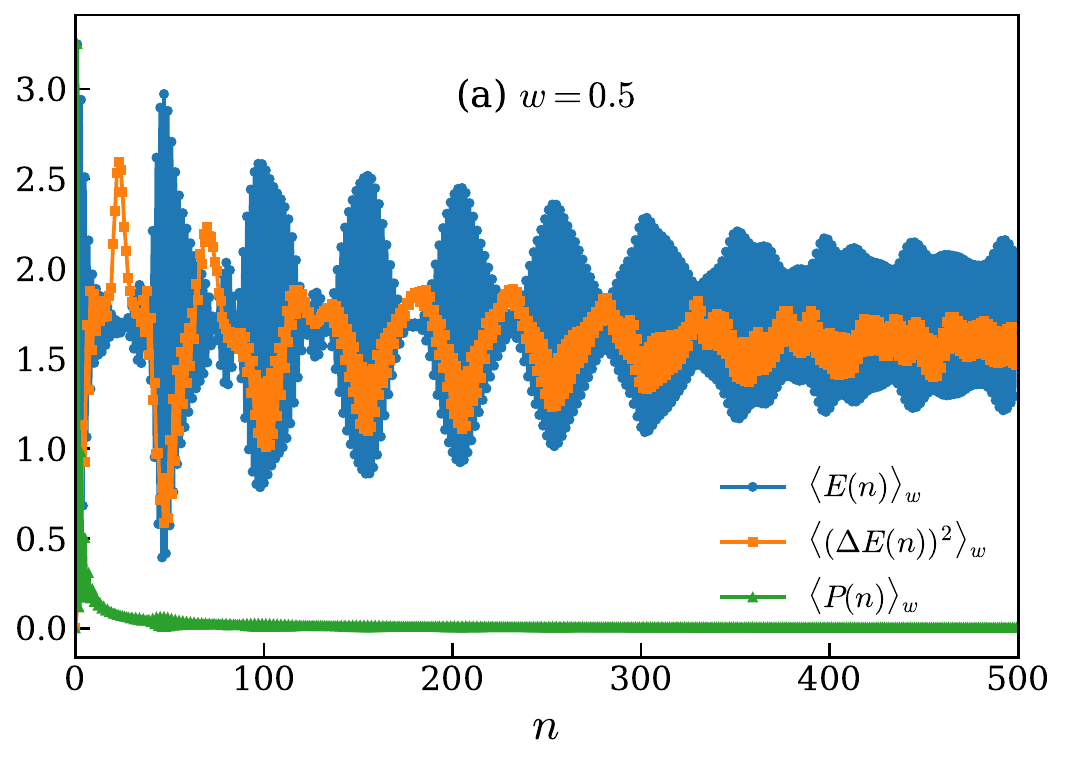}

\includegraphics[width=1.0\linewidth]%, height=0.8\linewidth]
{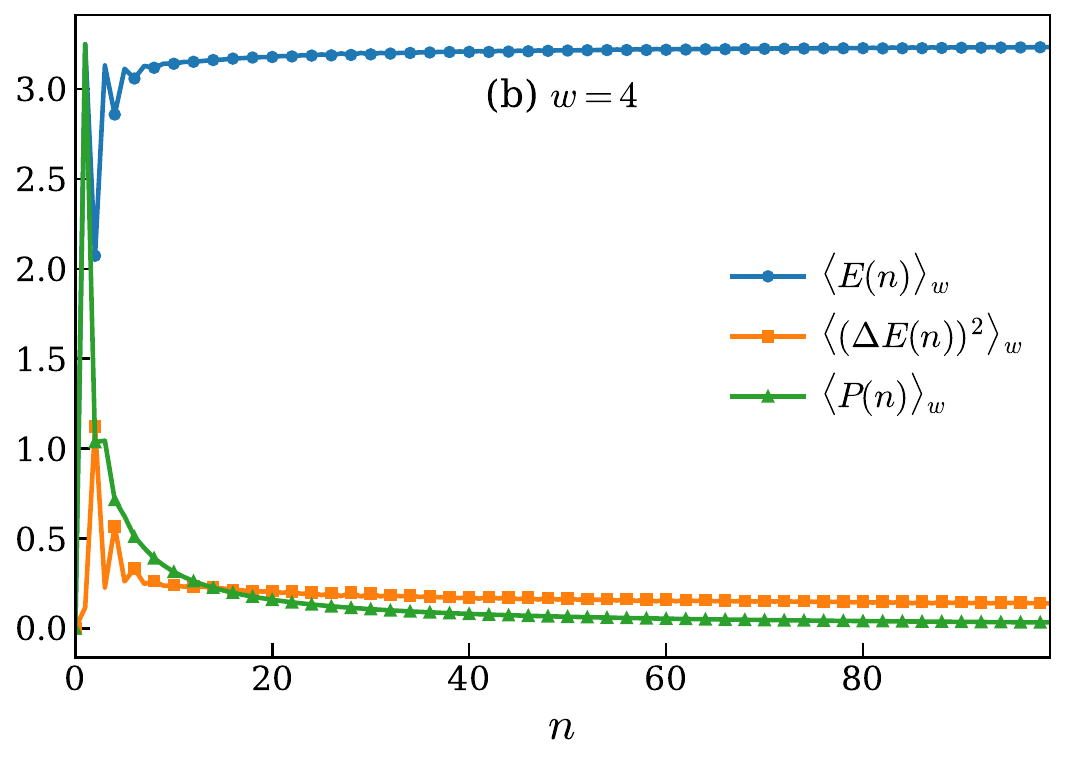}
\caption{Evolution of the disorder averaged stored energy, variance, and charging power as a function of the number of kicks $n$ for $N=14$, $p=\pi/2$, and disorder strengths (a) $w=0.5$ and (b) $w=4$. The battery is prepared in its ground state $\ket{\pi/2,0}$.} %\lnk{The inset in (a) shows the evolution over the first few kicks, highlighting the period-two oscillations of the stored energy.}}
\label{EVP_N14}
\end{figure}

\begin{figure}[h]
\centering
\includegraphics[width=1.0\linewidth]
{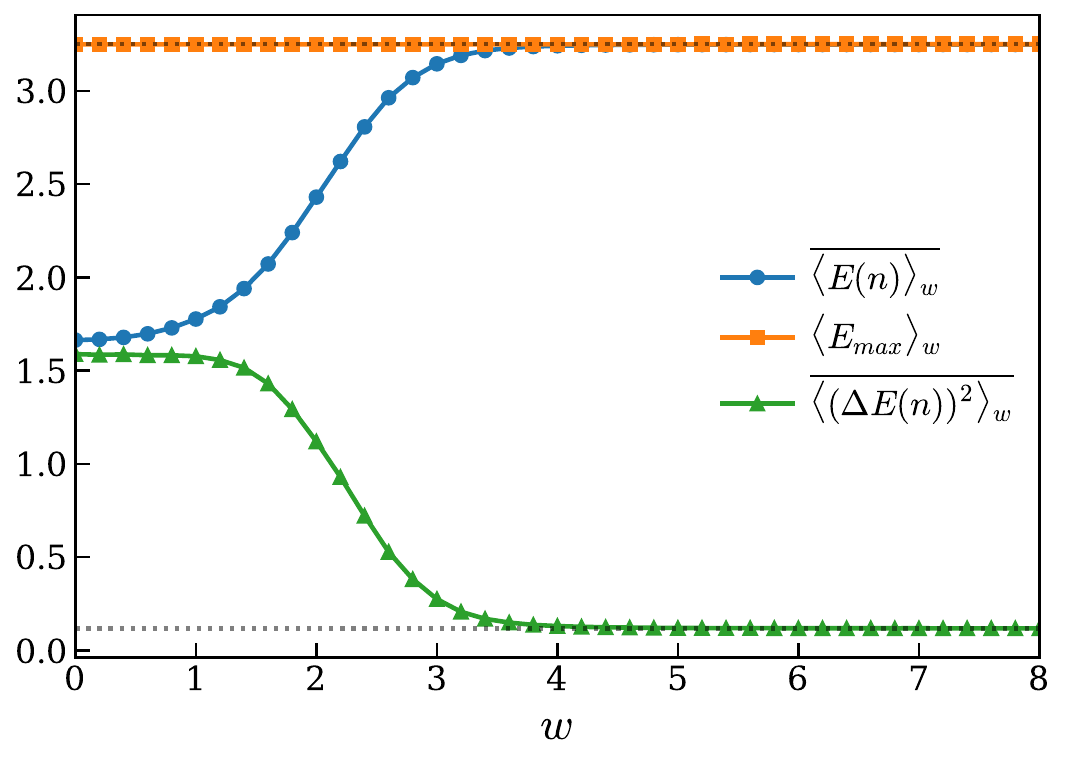}
\caption{Disorder and long time averaged stored energy and its variance, along with the maximum stored energy, shown as functions of the disorder strength $w$ for $k=1$, $p=\pi/2$, $N=14$. The battery is initialized in the spin-coherent state $\ket{\theta,\phi}=\ket{\frac{\pi}{2},0}$, corresponding to the ground state. The black dotted horizontal line indicates the random-matrix-theory (RMT) prediction of the variance evaluated in the full $2^N$-dimensional Hilbert space. With increasing disorder, we observe increased energy storage together with reduced energy fluctuations which is optimal for battery operation. }
\label{EV_N14_k1_gnd}
\end{figure}

We now discuss the numerical results obtained for the above battery cycle. 
Since the best performance of the battery in terms of energy storage is obtained when the battery is prepared in its ground state, we initialize the spins in the coherent state $\ket{\frac{\pi}{2},0}=\ket{+x}^{\otimes N}$. We first study the $k=1$ case, where the above spin coherent state is a period$-4$ orbit in the classical phase space since the corresponding phase-space point revisits its initial position after every four kicks.  
Using the charging protocol introduced earlier, we analyze the battery dynamics as the disorder strength $w$ and the number of kicks $n$ are varied.  
In Fig. \ref{EVP_N14}, we plot the variation of different quantities as a function of $n$ for $w=0.5$ (top) and $w=4$ (bottom). 
Clearly, when the disorder strength increases, the oscillations are increasingly suppressed and the steady state is reached in fewer kicks. We also note that the power peaks at $n=1$, so that maximum   power occurs at the first kick, and is therefore equal to energy at the first kick. %Since the maximum energy is deposited during the first kick (see Appendix \ref{appen:storedenergy}), the maximum power coincides with the maximum stored energy and does not reveal additional dynamical properties. 

We plot the long time averaged saturation value as a function of strength of disorder in Fig. \ref{EV_N14_k1_gnd}. We find that $\overline{\langle{E(n)}\rangle_{w}}$ increases monotonically with disorder and eventually saturates to its maximum value. This maximum saturation value coincides with the prediction from the random matrix theory in $2^N$ dimensions given by:
\begin{equation}
    -\left(\frac{k}{N}\right)\braket{J_x^2}_{RMT}-\left(-\frac{k}{N}\braket{J_x^2(t=0)}\right)=\frac{k}{4}(N-1),
\end{equation}
where $\braket{J_x^2}_{RMT}$ is its RMT value in FHS given by ${N}/{4}$ and $\braket{J_x^2(t=0)}=\frac{N^2}{4}$, evaluated at time $t=0$. See Appendix \ref{appen:RMT} for all RMT related calculations. Although the final steady-state energy converges to the same value $-\frac{k}{4}$ (independent of $N$), the initial-state energy depends on the size of the system, leading to an overall $N$-dependent offset in the stored energy. At the same time, $\overline{\langle{ (\Delta E(n))^2}\rangle_{w}}$ drops sharply with disorder and converges to its RMT value given by:
\begin{equation}
    \left(\frac{k}{N}\right)^2(\De{J_x^2})^2_{RMT}=\left(\frac{k}{N}\right)^2\frac{N(N-1)}{8},
\end{equation}
where $(\De{J_x^2})^2_{RMT}=\braket{J_x^4}_{RMT}-\braket{J_x^2}^2_{RMT}$ is its RMT value in FHS given by $N(N-1)/8$ signaling equilibration towards an effectively random state in the FHS (also derived in Appendix \ref{appen:RMT}). This behavior is indicative of a crossover from dynamics confined to the regular, permutation-symmetric subspace to dynamics that explore the full Hilbert space in a chaotic manner, consistent with earlier findings. These results highlight a favorable operating regime of the quantum battery, where maximal energy storage is achieved simultaneously with suppressed energy fluctuations. The performance of the battery is improved in the chaotic, permutation-symmetry-broken regime, where the dynamics extend over the full Hilbert space.

\begin{figure}[h]
\centering
\includegraphics[width=1.0\linewidth]%, height=0.8\linewidth]
{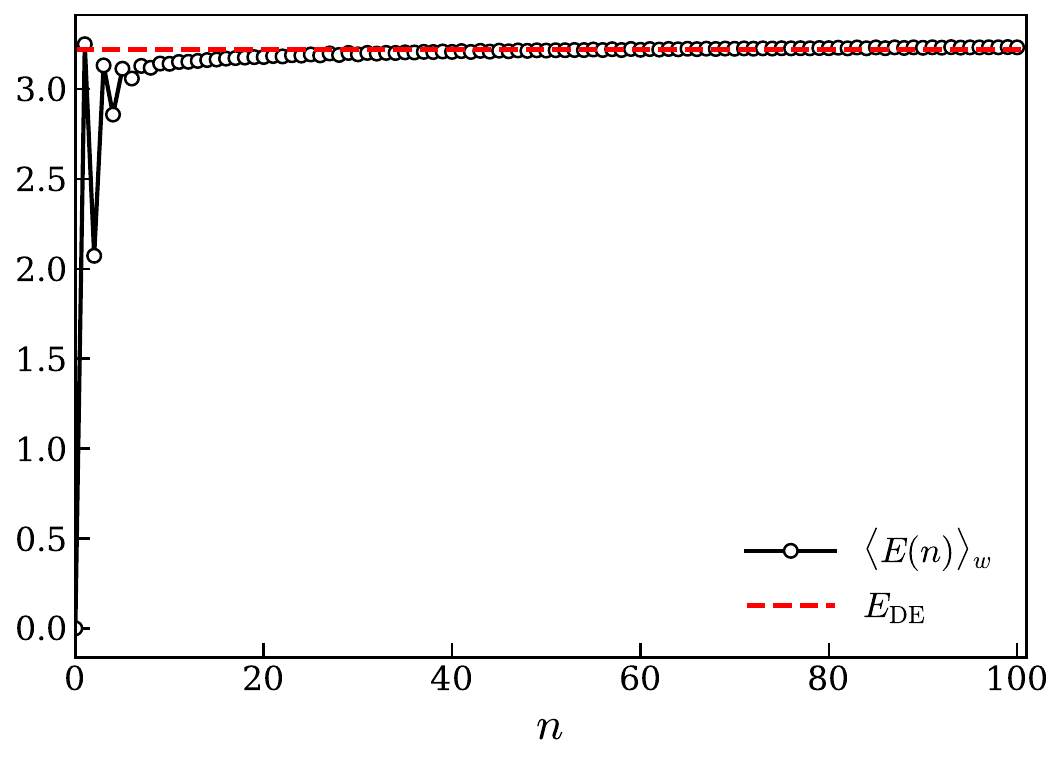}
\caption{Time evolution of the stored energy for 100 disorder realizations for $N=14$ and $w=4$ (starting from the ground state). The red dashed line indicates the diagonal-ensemble value ($E_{DE}=\sum_{\alpha} \vert{c_{\alpha}}\vert ^2\langle{\phi_{\alpha}}\vert{H_B}\vert{\phi_{\alpha}}\rangle- E_i$) where $E_i$ is the initial energy, which matches the RMT prediction, illustrating Floquet dephasing.  }
\label{diagonal_w4_N14}
\end{figure}

 In the Floquet eigen basis, the expectation value of an observable contains both diagonal and off-diagonal contributions. During the evolution, the off-diagonal terms acquire phase factors associated with quasi energy differences. Since the permutation symmetry is progressively broken as the disorder strength $w$ increases, a large number of Floquet states participate in the dynamics leading to an increasingly irregular quasienergy spectrum and enhanced phase scrambling among different Floquet components. Consequently, the off-diagonal contributions get suppressed, and the long-time dynamics is governed predominantly by the diagonal terms. The corresponding steady-state expectation value of the operator $\hat{O}$ is therefore described by the Floquet diagonal ensemble \cite{PhysRevX.4.041048, lazarides2014equilibrium},
\begin{equation}\label{eq_steady}
    \langle{\hat{O}}\rangle_{steady}=\sum_{\alpha} \vert{c_{\alpha}}\vert ^2\langle{\phi_{\alpha}}\vert{\hat{O}}\vert{\phi_{\alpha}}\rangle,
\end{equation}
where $\ket{\phi_{\alpha}}$ is the Floquet eigen state.
Consequently, the diagonal-ensemble value of the stored energy can be written as,
\begin{equation}\label{de}
E_{DE}=\sum_{\alpha} \vert{c_{\alpha}}\vert ^2\langle{\phi_{\alpha}}\vert{H_B}\vert{\phi_{\alpha}}\rangle- E_i, \end{equation} 
where $E_i$ is the initial energy. The steady-state energy stored agrees with the diagonal ensemble prediction given by Eq. \ref{de} reflecting Floquet dephasing as shown in Fig. \ref{diagonal_w4_N14}, which in turn coincides with the corresponding Random Matrix Theory (RMT) value, suggesting the emergence of an ergodic Floquet regime \cite{peres1984ergodicity}.

\begin{figure}[h]
\centering

\includegraphics[width=1.0\linewidth]%, height=0.8\linewidth]
{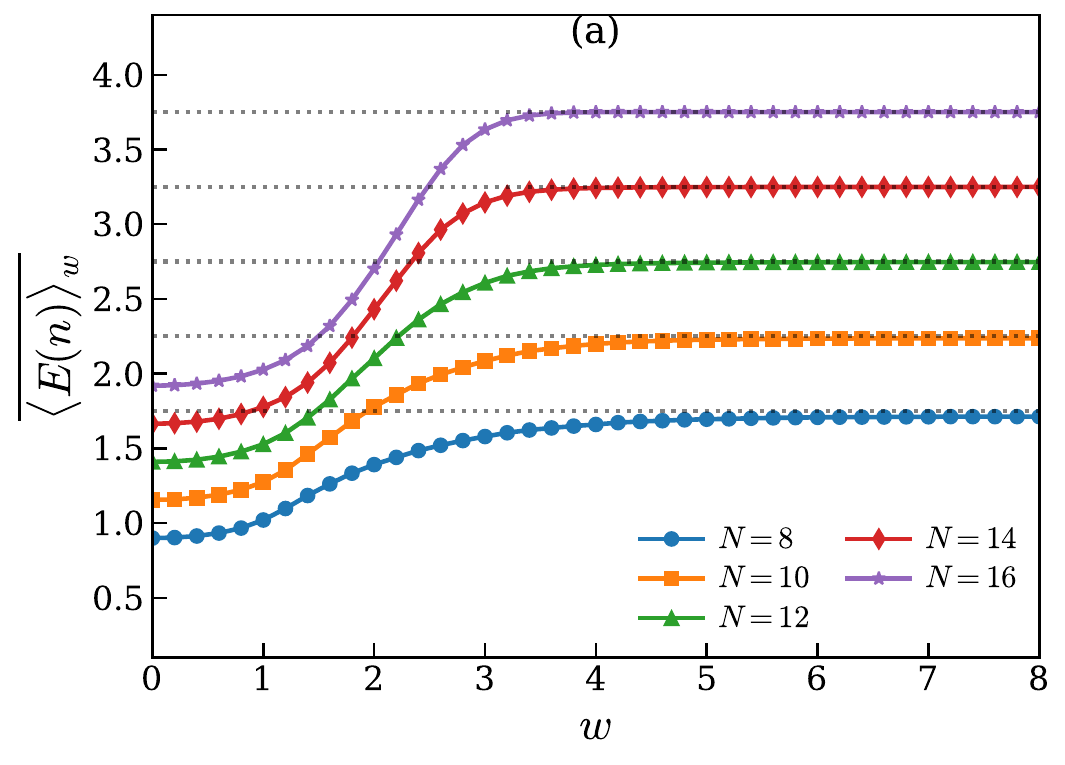}

\includegraphics[width=1.0\linewidth]%, height=0.8\linewidth]
{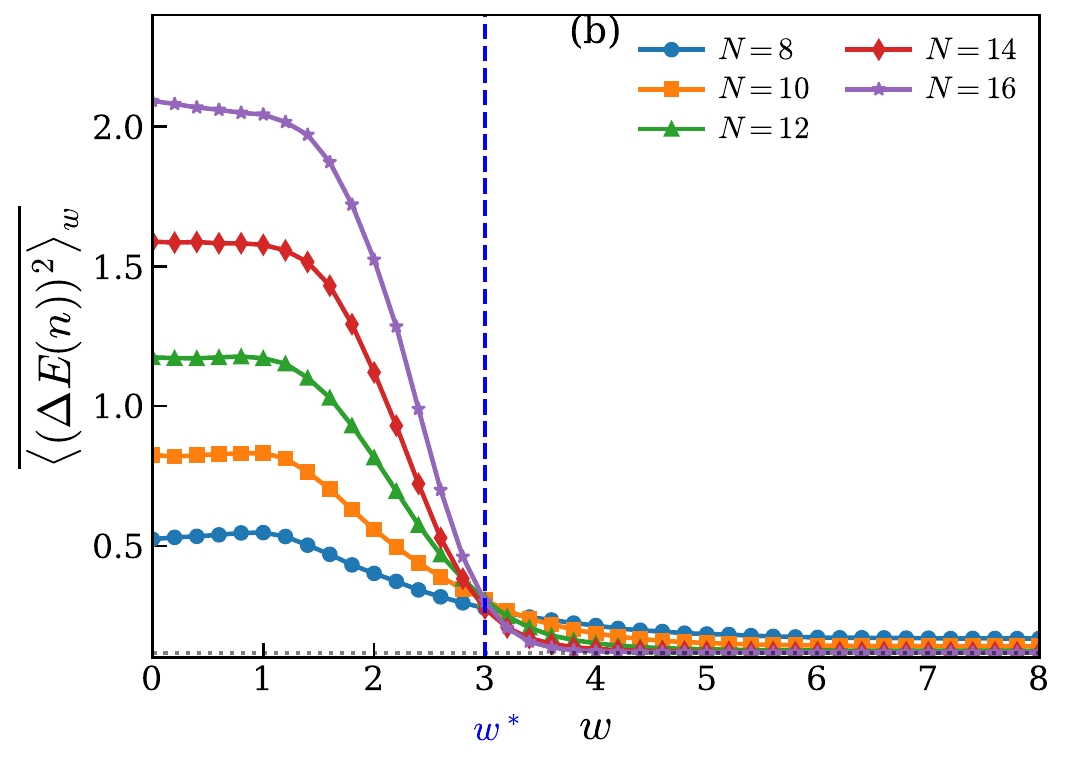}
\caption{Disorder and long time averaged (a) stored energy and (b) variance plotted as a function of the disorder strength $w$ for different system sizes $N$, with $k=1$, $p=\pi/2$. The battery is initialized in its ground-state $\ket{\theta,\phi}=\ket{\pi/2,0}$. The black horizontal dotted line in (a) and (b) denote the corresponding random-matrix-theory (RMT) prediction evaluated in the $2^N$ dimensional Hilbert space. The energy approaches the RMT value with increasing system size as seen in (a). The variance likewise converges towards the RMT prediction as shown in (b). In addition, the variance curves for different $N$ exhibit a crossing at a characteristic disorder strength $w^{*}\sim 3$. The vertical dashed line in blue separates the two different operating regimes, namely, incomplete dephasing and fully ergodic regime, of the battery.}
\label{allN}
\end{figure}

Our results are consistent across several system sizes as shown in Fig. \ref{allN}. We observe crossing of variance curves for different system sizes around the disorder strength $w^*$. While limited to finite system sizes, the convergence of the stored energy and its variance towards the RMT predictions suggests a crossover from regular to ergodic dynamics when $k$ is set unity \cite{PhysRevE.50.888,d2016quantum, peres1984ergodicity, feingold1984ergodicity, khaymovich2019eigenstate, mondaini2017eigenstate, rigol2008thermalization, PhysRevX.4.041048, lazarides2014equilibrium}. Similar results are also obtained for other initial states as well.
%Hence, the crossing point $w^*$ naturally emerges as a dynamical ergodic crossover point. 
We find the existence of $w^*$ such that, for $w<w^*$, the system exhibits incomplete dephasing and for $w>w^*$, the dynamics approaches a fully ergodic regime in the thermodynamic limit. Therefore, $w^*$ marks the onset of ergodicity. Notably, the variance provides a more sensitive diagnostic of this ergodic crossover than the stored energy.

\begin{figure}[h]
\centering
\includegraphics[width=1.0\linewidth]%, height=0.8\linewidth]
{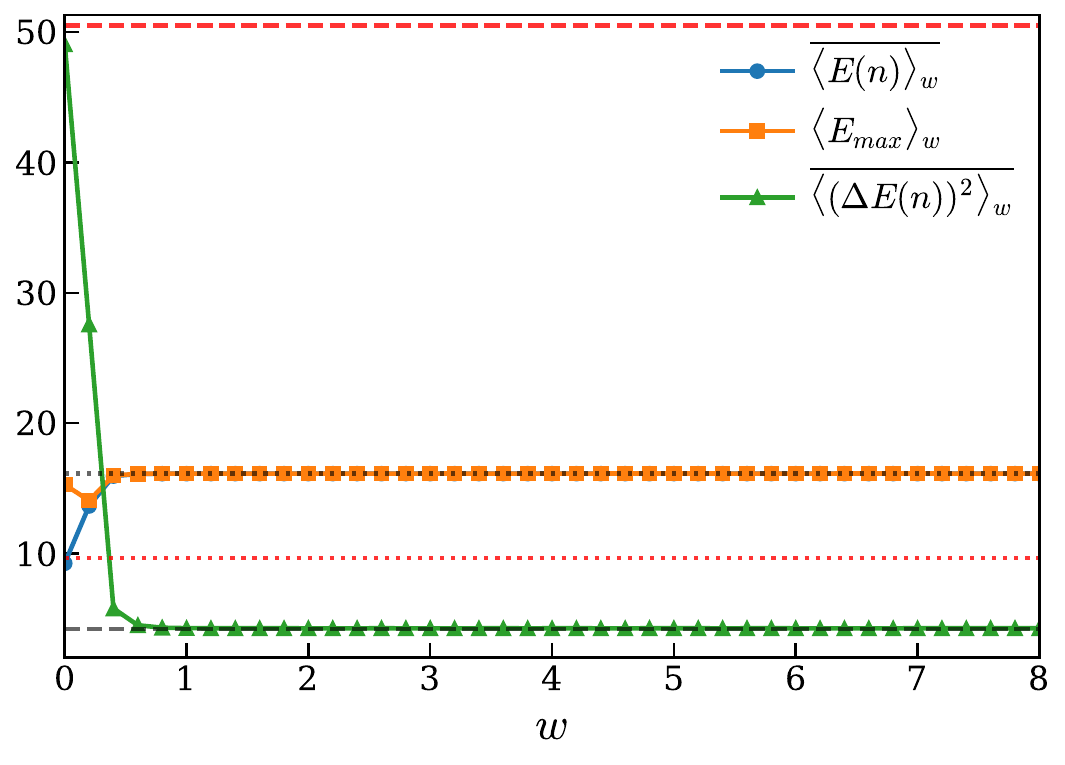}
\caption{Disorder and long time averaged stored energy and its variance, along with the maximum stored energy, shown as a function of the disorder strength $w$ for $k=6$, $p=\pi/2$, $N=14$, and a different initial state $\ket{2,0}$ to confirm the state independence of the results. The black (red) dotted line denote the RMT prediction of $\overline{\braket{E(n)}_w}$ evaluated in the full $2^N$-dimensional Hilbert space ($N+1$-dimensional permutation symmetric subspace). Similarly, the black (red) dashed line represents the energy variance calculated using RMT in full $2^N$ Hilbert space (N+1-dimensional space).}
\label{EV_N14_k1_scs}
\end{figure}

We also study the behavior when $k$ is set to $6$ with some other initial state, say a spin coherent state $\ket{2,0}$ that lies in the chaotic region of the corresponding classical phase space. Without disorder, the dynamics follow RMT predictions within the permutation-symmetric ($N+1$)-dimensional subspace. Upon introducing disorder, symmetry breaking drives the system toward RMT behavior in the full $2^N$-dimensional Hilbert space, as shown in the Fig. \ref{EV_N14_k1_scs}. We find that the stored energy increases while the variance decreases as the strength of disorder is increased. This behavior can be traced to the availability of a larger set of accessible energy eigen states compared to the symmetry restricted PSS, leading to more efficient energy storage similar to $k=1$ behavior. These results further reinforce our earlier conclusions, identifying the symmetry-broken regime with chaotic dynamics as the optimal operating regime of the battery within the parameter ranges studied.

\begin{figure}[h]
\centering
\includegraphics[width=1.0\linewidth]%, height=0.8\linewidth]
{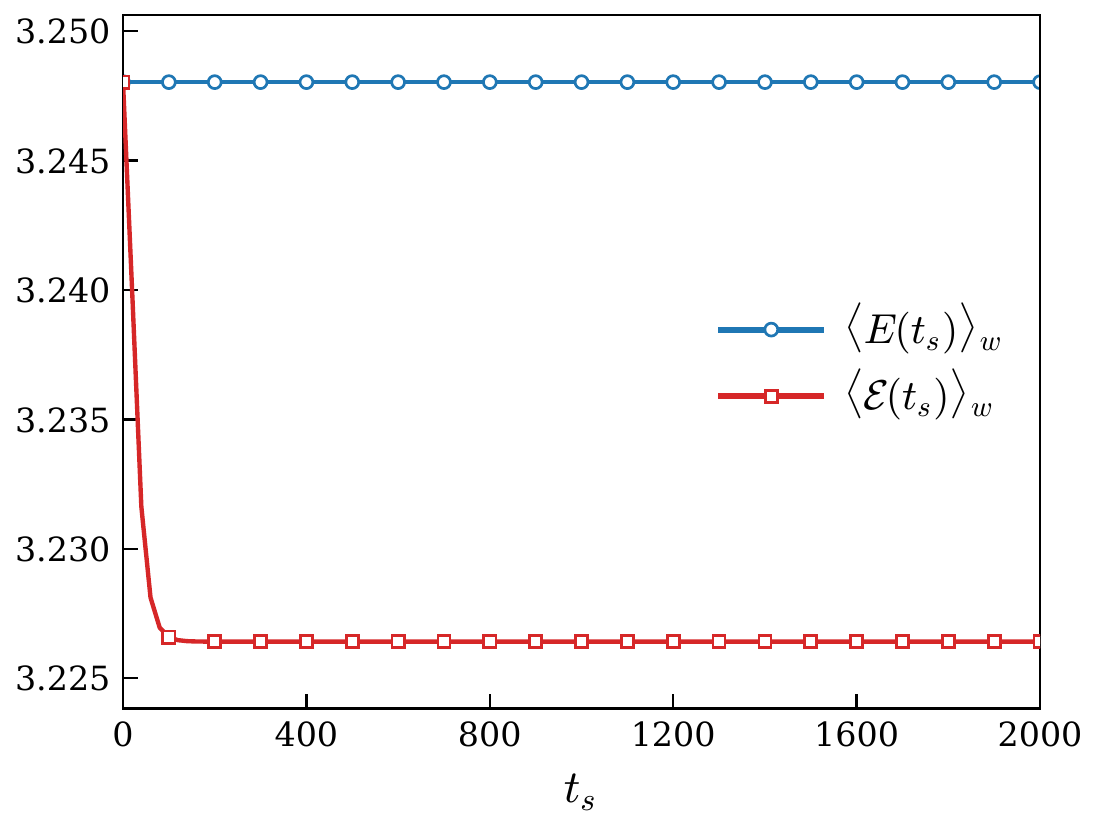}
\caption{Time evolution of ergotropy $\langle{\mathcal{E}(t_s)}\rangle_w$ and  stored energy $\braket{E(t_s)}_w$,  during the storage phase under energy dephasing noise (where $t_s$ is the storage time), with the storage dynamics initiated from the states obtained after 800 charging kicks and averaged over 100 disorder realizations. Parameters are $N=14$, $p=\pi/2$, $w=7$, $k=1$ and $\gamma=0.02$ with the battery system initially prepared in the $\ket{\frac{\pi}{2},0}$ ground state before charging it.}
\label{Storage14}
\end{figure}

\begin{figure}[h]
\centering
\includegraphics[width=1.0\linewidth]%, height=0.8\linewidth]
{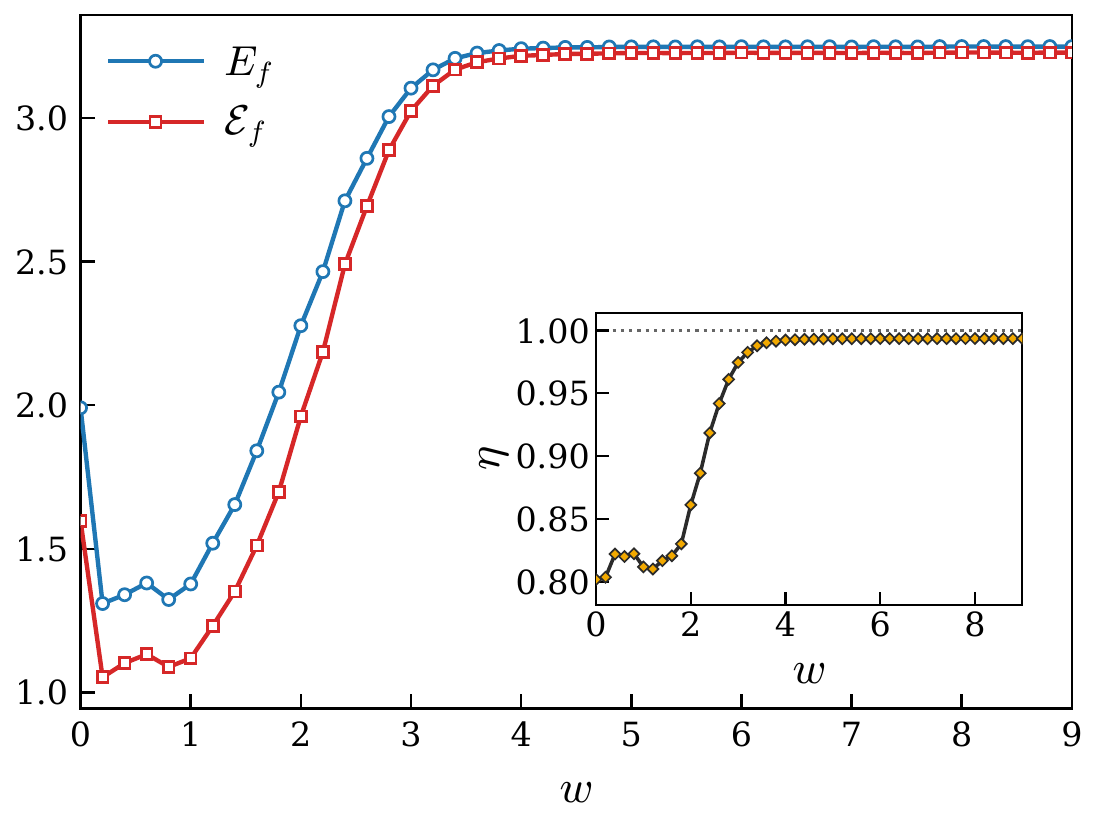}
\caption{Final ergotropy $\mathcal{E}_f$ and stored energy $E_f$ as a function of the disorder strength $w$ at the end of the storage phase for $N=14$, $p=\pi/2$, $k=1$ and $\gamma=0.02$. The inset shows the efficiency, $\eta$, as a function of $w$, with the black dotted line indicating $\eta=1$.}
\label{storage_w}
\end{figure}

We now present and discuss the results obtained for the storage and extraction stages. Here, for each disorder realization, we consider the state obtained after 800 charging kicks (sufficient for the system to reach the steady state for large disorder strengths) as the initial state for the subsequent storage dynamics discussed in Sec.\ref{sec:storage}. This involves evolving the initial state under the Lindblad master equation given in Eq. \ref{lindblad} up to time $t_s=2000$, by which the ergotropy, discussed next, reaches its saturation value. The ergotropy is then evaluated separately for each final state and subsequently disorder averaged. During the storage dynamics, we find that the stored energy remains constant, as expected for the pure-dephasing channel considered here. In contrast, the ergotropy exhibits a weak decay as shown in Fig. \ref{Storage14}. %Repeating the procedure for a range of disorder strengths $w$, we evaluate the final ergotropy and stored energy at the end of the storage interval. 
In Fig. \ref{storage_w}, we plot the final ergotropy ${\mathcal{E}_f}$, and final stored energy $E_f$, as a function of $w$, along with the corresponding efficiency, defined as  $\eta=\mathcal{E}_f/E_f$ (see inset). With increasing disorder, the ergotropy gets closer to the stored energy, resulting to an efficiency $\eta \simeq 1$. It is surprising that even after a noisy storage dynamics, a large fraction of the stored energy remains extractable as useful work in the battery. Remarkably, this high-efficiency regime coincides with the regime of enhanced charging performance, further highlighting the relevance of the strongly disordered, permutation-symmetry-broken chaotic regime for battery operation. The numerical results discussed above, starting from the charging, storage and extraction is summarized in the schematic given in Fig. \ref{battery cycle}.

%We consider the disorder-averaged battery state obtained after charging for 800 Floquet kicks(sufficient for the system to reach the steady state for sufficiently large disorder strengths) as discussed in Sec.\ref{sec:storage}. This state is used as the initial condition for the subsequent open-system dynamics as part of the storage where it is evolved in the presence of dephasing noise governed by Lindblad dynamics. The noisy evolution (Eq. \ref{lindblad}) is continued up to 2000 time steps by which the ergotropy has already saturated, after which the ergotropy of the final state is evaluated. It should be noted that the battery retains a nonzero ergotropy even though the stored energy approaches its equilibrium value, reflecting the fact that equilibration of observable expectation values does not necessarily imply passivity of the equilibrated battery state, allowing a finite amount of extractable work to survive even at long times.  While the stored energy remains constant under pure dephasing, the ergotropy decays slightly as shown in Fig. \ref{Dephase12}, demonstrating that decoherence primarily limits usefulness rather than capacity. Even after including a realistic storage stage modeled by pure dephasing, the higher-disorder regime retains a larger extractable work (ergotropy) than the lower-disorder regime as shown in Fig. \ref{Ergotropy}. For the parameters considered, higher disorder during charging leads to a larger residual ergotropy after storage compared to lower disorder.

\section{SUMMARY AND DISCUSSION}
\label{sec:summary}

Previous studies have shown that adding disorder to the kicked top model gives rise to distinct dynamical regimes: a permutation-symmetric regime and a full-Hilbert-space chaotic regime. In the present work, we exploit these different regimes during the charging phase of a  battery consisting of all-to-all interacting spin chains which is charged by introducing disorder in the interaction direction and kicking in the transverse direction. We find that the battery operates optimally in the large disorder limit when the dynamics explore the full Hilbert space, where observables such as the stored energy and energy variance of the battery converge towards their random-matrix-theory predictions, with the variance serving as a more sensitive indicator of this crossover. Our results also indicate a reliable battery operating regime at large disorder strengths $w$, where the stored energy is large, the variance of the energy is minimum and the ergotropy constitutes nearly the entire stored energy resulting in an efficiency close to unity. Although our analysis is restricted to finite system sizes accessible to numerical simulations, the observed trends are robust across the various system sizes studied. 

While we focussed on a disorder free all-to-all interacting battery, one can also study other types of battery with similar charging scheme. We find the battery model studied in this work to be the most efficient. For example, a non-interacting battery does not exhibit a clear distinction between the permutation-symmetric and symmetry-broken regimes, since the relevant spin expectation values vanish in both cases, leading to qualitatively similar energy behaviour for all values of $w$. On the other hand, introducing disorder directly into the battery Hamiltonian causes the energy variance to depend on disorder strength explicitly, and to grow monotonically with increasing disorder (see Appendix \ref{appen:batterydisorder} for details), which is not preferred for battery stability. Therefore, the clean, interacting battery provides a balanced and effective choice: it achieves large energy storage while maintaining controlled fluctuations and permits a clear separation between symmetry-restricted and symmetry-broken dynamical regimes. With this choice of battery and the adopted charging protocol, we systematically probed how disorder in the charging dynamics enhances the operational characteristics of the battery. We complete the battery cycle by studying a storage stage with pure dephasing noise followed by energy extraction, which also shows an advantage in the large-$w$ limit.

We believe that the proposed quantum battery and charging protocol can be physically realized in trapped-ion platforms, which offer remarkable control over collective spin interactions and global rotations\cite{milburn1999simulating,Monroe2021,Xu2010,Sieberer2019,PhysRevLett.82.1835,PhysRevA.62.022311,Shapira2025,Lu2025}.

Our results indicate the importance of integrating chaos, ergodicity, and symmetry constraints in the development of efficient quantum batteries. An important direction for future work is to investigate the impact of open-system dynamics during the charging phase itself to assess the robustness of the ergodic regime under pure dephasing noise.

\begin{acknowledgments}
UD acknowledges support from Anusandhan National Research
Foundation (ANRF), Government of India, through
grant No. SPG/2022/000708. LNK acknowledges the HPC facilities, Chandra and Madhava, at IIT Palakkad where the computations were carried out. The authors also thank Manju C for insightful discussions. 
\end{acknowledgments}
\appendix

\section{Pure dephasing noise}
\label{appen:noise}

Here, we detail the pure-dephasing model considered during the storage process, implemented through a Lindblad dissipator\cite{Shastri2025} generated by the battery Hamiltonian. The corresponding Lindblad operator is chosen as
\begin{equation}
    \mathcal{L}= \sqrt{2\gamma} H_B,
\end{equation}
where $H_B$ is the battery Hamiltonian. 
Then the dissipator is as follows:
\begin{eqnarray}
    \mathcal{D}(\rho)&=&\mathcal{L}\rho\mathcal{L}^{\dagger}-\frac{1}{2}\{\mathcal{L}^{\dagger}\mathcal{L},\rho \}  \nonumber \\
   &=& -\gamma [H_B, [H_B, \rho]].
\end{eqnarray}
The corresponding master equation is:
\begin{equation}\label{open}
    \frac{d \rho}{ d t}= -i[H_B, \rho]-\gamma [H_B, [H_B, \rho]].
\end{equation}
To understand the action of the dissipator, we evaluate its matrix elements in the energy eigenbasis of the battery Hamiltonian, $H_B|n\rangle = E_n |n\rangle$. The dissipative term takes the form
\begin{equation}
(\mathcal{D}(\rho))_{mn}=-\gamma (E_m-E_n)^2 \rho_{mn}.
\end{equation}
It follows immediately that the diagonal elements remain unaffected,
\begin{equation}
(\mathcal{D}(\rho))_{nn}=0,
\end{equation}
so that the populations in the energy basis are conserved. In contrast, the off-diagonal elements evolve as
\begin{equation}
\rho_{mn}(t)=\rho_{mn}(0)e^{-\gamma (E_m-E_n)^2 t}, (m\neq n),
\end{equation}
and therefore decay exponentially with a rate proportional to the squared energy difference. Consequently, the dissipator suppresses coherences while leaving the populations unchanged, which is the hallmark of pure dephasing. Therefore, the elements of the density matrix evolve under the open dynamics given by Eq. \ref{open} as 
\begin{equation}
\rho_{mn}(t)=\rho_{mn}(0)e^{-i(E_m-E_n)t}e^{-\gamma (E_m-E_n)^2 t}.
\end{equation}

\section{Random-Matrix-Theory predictions}
\label{appen:RMT}

Here, we derive the RMT expectation values of the relevant
observables by averaging over Haar-random pure states. For a Haar-random pure state ($\vert{\psi}\rangle$) in a ($D$)-dimensional Hilbert space, the ensemble average of an observable ($A^k$) is given by \cite{haake1991quantum,lakshminarayan2025chaos}
\begin{equation}
\langle A^k \rangle_{\mathrm{RMT}}
=
\frac{1}{D}\mathrm{Tr}\left(A^k\right),
\end{equation}
where (D) is the Hilbert-space dimension.

Below, we first calculate the RMT benchmarks for $\braket{J_x^2}$ and $\braket{J_x^4}$  within the ($N+1$)-dimensional permutation-symmetric subspace, followed by those for the full Hilbert space. These results are subsequently used to evaluate the stored energy and its variance.

\subsection{Permutation-Symmetric (N+1)-Dimensional Hilbert Space}

For calculating various quantities in the permutation-symmetric case (see Fig. \ref{EV_N14_k1_scs}), it is convenient to work in the $\ket{j,m}$ basis, where $m$ denotes the magnetization along the $x$-direction. Setting $j=\frac{N}{2}$, we obtain 
\begin{equation}
    \langle{J_x^2}\rangle = \sum_{m=-j}^{+j}m^2 = \frac{N(N+1)(N+2)}{12},
\end{equation}
resulting in  
\begin{equation}
     \langle{J_x^2}\rangle_{RMT}=\frac{N(N+2)}{12}.
\end{equation}
Similarly, we get 
\begin{equation}
    \langle{J_x^4}\rangle = \sum_{m=-j}^{+j}m^4 = \frac{N(N+1)(N+2)(3N^2+6N-4)}{240},
\end{equation}
which gives 
\begin{equation}
     \langle{J_x^4}\rangle_{RMT} = \frac{N(N+2)(3N^2+6N-4)}{240}.
\end{equation}

\begin{figure}
\centering
\includegraphics[width=1.0\linewidth] %height=0.8\linewidth]
{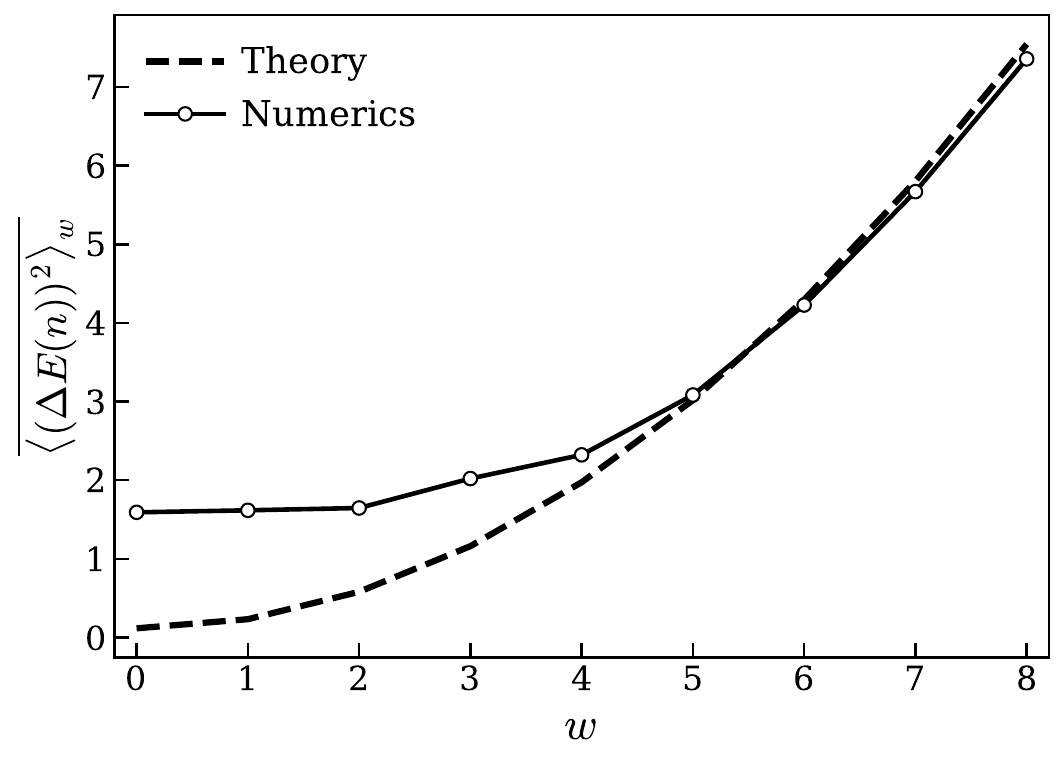}
\caption{Comparison of the theoretically predicted energy variance with numerically obtained values for a disordered quantum battery, demonstrating close agreement  for larger disorder strengths $w$ for system size $N=14$, initial state prepared as the ground state at $(\pi/2,0)$ and kick strength $p=\pi/2$.}
\label{Analytical12}
\end{figure}

\subsection{Full $2^N$ Hilbert space}

We repeat the calculations for the full Hilbert space. Note that $\text{Tr}(J_x^2)=\frac{1}{4}\text{Tr}(\sum_{i,j=1}^{N}\sigma_i^x \sigma_j^x)$ gives a nonzero trace only if $i=j$ ($\sigma_x^2=I$). Therefore, $\text{Tr}(J_x^2)=\frac{1}{4} \times N \times 2^N$ resulting to 
 \begin{equation}
     \langle{J_x^2}\rangle_{RMT}=\frac{\text{Tr}(J_x^2)}{D}=\frac{N}{4}.
 \end{equation}
Similarly, in order to have a non-zero value for $\text{Tr}(J_x^4)=\frac{1}{16}\text{Tr}(\sum_{i,j,k,l=1}^{N}\sigma_i^x \sigma_j^x \sigma_k^x \sigma_l^x)$, there are two possibilities. (i) If $i=j=k=l$, $\text{Tr}(I_{2^N \times 2^N})=2^N$ contributes $N \times 2^N$ to the trace. (ii) Index configurations of the form $(a,a,b,b)$ with $a\neq b$ involving only two distinct indices. There are 6 distinct permutations of such an index arrangement. In addition, there are $\frac{N(N-1)}{2}$ possible choices for the pair $(a,b)$. Consequently, these terms contribute a total of $6 \times \frac{N(N-1)}{2} \times 2^N$ to the trace. Hence, \begin{equation}
\text{Tr}(J_x^4)=\frac{1}{16} \times 2^N \left(N+6 \frac{N(N-1)}{2}\right)=2^N\left(\frac{3N^2-2N}{16}\right),  \end{equation}
so that
\begin{equation}
     \langle{J_x^4}\rangle_{RMT}=\frac{3N^2-2N}{16},  
\end{equation}
and its variance becomes:
\begin{equation}
    (\Delta J_x^2)_{RMT}=\frac{N(N-1)}{8}.   
\end{equation}

\section{Battery with disorder}
\label{appen:batterydisorder}

While the main text considers a clean battery, this appendix examines the effect of introducing disorder into the battery Hamiltonian under the same charging protocol, consisting of a sequence of kicks during the charging stage. We find that the energy variance scales with the disorder strength, which is an undesirable effect for the battery.

The battery Hamiltonian is chosen to be
\begin{equation}
    H_{D}=-\frac{k}{2N }\sum_{l<l^{'}}(1+\epsilon_{ll^{'}})\sigma_{l}^{x}\sigma_{l^{'}}^{x}.
\end{equation}
The evolution is again described by Eq. \ref{dynamics1}. Within the framework of random matrix theory (RMT), the energy variance is predicted to take the form:
\begin{equation}
    \text{Var}(H_{D})=\frac{1}{2^{N}}\text{Tr}(H_{D}^2)-\left(\frac{1}{2^N}\text{Tr}(H_{D})\right)^2.
\end{equation}
For any $l \neq l^{'}$, $\text{Tr}(\sigma_{l}^{x}\sigma_{l^{'}}^{x})=0$, yielding 
\begin{equation}\label{VarHd}
    \text{Var}(H_{D})=\frac{1}{2^{N}}\text{Tr}(H_{D}^2),
\end{equation}
where
\begin{equation}
    H_{D}^2=\left(\frac{k}{2N}\right)^2\sum_{l<l^{'}}\sum_{m<m^{'}}(1+\epsilon_{ll^{'}})(1+\epsilon_{mm^{'}})\sigma_{l}^{x}\sigma_{l^{'}}^{x}\sigma_{m}^{x}\sigma_{m^{'}}^{x}.
\end{equation}
For $(l,l^{'})=(m,m^{'})$, \begin{equation} \text{Tr}(\sigma_{l}^{x}\sigma_{l^{'}}^{x}\sigma_{m}^{x}\sigma_{m^{'}}^{x})=\text{Tr}((\sigma_{l}^{x}\sigma_{l^{'}}^{x})^2)=\text{Tr}(I)=2^N,  \nonumber \end{equation} 
whereas for $(l,l^{'}) \neq (m,m^{'})$, $\text{Tr}(\sigma_{l}^{x}\sigma_{l^{'}}^{x}\sigma_{m}^{x}\sigma_{m^{'}}^{x})=0$, 
so that
\begin{equation}\label{HD2}
    \text{Tr}(H_D^2)=\left(\frac{k}{2N}\right)^2\sum_{l<l^{'}}(1+\epsilon_{ll^{'}})^2 2^N.
\end{equation}
The random variables are taken from a Gaussian distribution with zero mean and standard deviation $w$ so that  %$\sum_{l<l'}(1+\epsilon_{ll'})=1+w^2$. 
\begin{equation}
    \mathbb{E}\!\left[(1+\epsilon_{ll'})^2\right]=1+w^2.
\end{equation}

\begin{comment}    

\[
\epsilon_{ll'}\stackrel{\mathrm{i.i.d.}}{\sim}\mathcal{N}(0,w^2),
\]
we have
\[
\mathbb{E}[\epsilon_{ll'}]=0,
\qquad
\mathbb{E}[\epsilon_{ll'}^2]=w^2.
\]
Therefore,
\[
\mathbb{E}\!\left[(1+\epsilon_{ll'})^2\right]=1+w^2.
\]
\end{comment}
Noting that there are $N(N-1)/2$ interaction pairs, Eq. \ref{VarHd} now reduces to
\begin{equation}\label{final}
\text{Var}(H_D)
=
\frac{k^2}{8}\frac{N-1}{N}\,(1+w^2).
\end{equation}
As shown in Fig. \ref{Analytical12}, the analytical expression Eq. \ref{final}  is in close agreement with the numerical result for large $w$, as expected.

%\newpage
%\bibliographystyle{apsrev4-2}
\bibliography{references}

\end{document}